\documentstyle[11pt,amsfonts]{article}
\renewcommand{\theequation}{\arabic{equation}}
\newcommand{\be}{\begin{equation}}
\newcommand{\ee}{\end{equation}}
\newcommand{\bea}{\begin{array}}
\newcommand{\ea}{\end{array}}
\newcommand{\beqa}{\begin{eqnarray}}
\newcommand{\eeqa}{\end{eqnarray}}
\newcommand{\bean}{\begin{eqnarray*}}
\newcommand{\eean}{\end{eqnarray*}}

\def\up#1{\leavevmode \raise.16ex\hbox{#1}}

\newcommand{\gapproxeq}{\lower
 .7ex\hbox{$\;\stackrel{\textstyle >}{\sim}\;$}}
\newcommand{\lapproxeq}{\lower .7ex\hbox{$\;\stackrel
{\textstyle <}{\sim}\;$}}

\renewcommand{\theequation}{\thesection.\arabic{equation}}

\newcounter{appendice}
\newcommand{\appendice}
{
\setcounter{equation}{0}
\renewcommand{\theequation}{\Alph{appendice}.\arabic{equation}}
\addtocounter{appendice}{1}
{\bf Appendix \Alph{appendice}}
}

\def\thebibliography#1{{\bf REFERENCES\markboth
 {REFERENCES}{REFERENCES}}\list
 {[\arabic{enumi}]}{\settowidth\labelwidth{[#1]}\leftmargin\labelwidth
 \advance\leftmargin\labelsep
 \usecounter{enumi}}
 \def\newblock{\hskip .11em plus .33em minus -.07em}
 \sloppy
 \sfcode`\.=1000\relax}

\begin{document}


\centerline{ \LARGE  Particle-like solutions to classical noncommutative gauge theory} 

\vskip 2cm

\centerline{ {\sc A. Stern}  }

\vskip 1cm
\begin{center}
Department of Physics, University of Alabama,\\
Tuscaloosa, Alabama 35487, USA
\end{center}

\vskip 2cm

\vspace*{5mm}

\normalsize
\centerline{\bf ABSTRACT} 
 
\vspace*{5mm}

We  construct perturbative static solutions to the classical field equations of noncommutative $U(1)$ gauge theory   for the three cases: a)  space-time noncommutativity, b) space-space noncommutativity and  c) both a) and b).  The solutions tend to the Coulomb solution at spatial infinity and are valid for intermediate values of the radial coordinate $r$.  They yield a  self-charge inside a sphere of radius $r$ centered about the origin which increases with decreasing $r$ for case a), and decreases  with decreasing $r$ for case b).   For case a) this may mean that the exact solution screens an infinite charge at the origin, while for case b) it is plausible that the charge density is well behaved at the origin, as happens in Born-Infeld electrodynamics.  For both cases a) and b)  the self-energy in the intermediate region  grows faster as $r$ tends to the origin  than that of the Coulomb solution.  It then appears that the  divergence of the classical self-energy  is more severe in the noncommutative theory than it is in the corresponding commutative theory.  
We  compute the lowest order effects of these solutions  on the hydrogen atom spectrum  and use them to put experimental bounds on the space-time and space-space
noncommutative scales. For the former, we get a significant improvement over previous bounds. We find that cases a) and  b) have different experimental signatures.

\newpage
\scrollmode

\section{Introduction}

\setcounter{equation}{0}

Recent  quantum field theory investigations  of noncommutative gauge theories have not been straightforward due to UV/IR mixing and related problems of renormalization. (For reviews, see \cite{Douglas:2001ba},\cite{Szabo:2001kg}.)
  On the other hand, important progress has been made in  understanding the renormalization of other noncommutative field theories, in particular   the noncommutative $\phi^4$-model \cite{Grosse:2004yu}.  There has also been a very recent attempt to apply similar methods to noncommutative gauge theories \cite{Blaschke:2008yj}.  Concerning UV-IR mixing, it is absent from some theories upon using an approach based on the twisted action of the Poincare group \cite{Balachandran:2005pn},\cite{Balachandran:2008gr}, and it can be understood for noncommutative gauge theories in terms of an induced gravity action \cite{Steinacker:2007dq},\cite{Grosse:2008xr}.
 Nevertheless, the renormalization of noncommutative gauge theories is  yet to be fully understood. 
  
    In light of the difficulties with the quantum field theory it may be useful to have a closer look at the classical field theory, and more specifically, for some behavior which may provide clues to these difficulties. 
Classical aspects of noncommutative gauge theories, and in particular their solutions, have been of recent interest.  There has been much work done on the noncommutative analogues of vortex, monopole and instanton solutions\footnote{ For a small sample, see \cite{Hashimoto:1999zw},\cite{Bak:1999id},\cite{Nekrasov:2000ih},\cite{Lozano:2000qf},\cite{Martin:2005vr},\cite{Stern:2006zt}.}, as well as the solutions of  general relativity.\footnote{  Many different noncommutative deformations of general relativity have been found \cite{Chamseddine:2000si},\cite{Aschieri:2005yw},\cite{Calmet:2006iz},\cite{Banerjee:2007th}.
 They have been used to compute  corrections to black hole solutions.  Approaches have ranged from smearing out the point singularity\cite{Nicolini:2005vd}, to perturbatively solving  noncommutative analogues of general relativity\cite{Chaichian:2007we},\cite{Mukherjee:2007fa},\cite{Kobakhidze:2007jn}.   Corrections to conical singularities have also been computed \cite{Pinzul:2005ta}.}
 
    Here we shall be concerned with classical aspects of noncommutative $U(1)$ gauge theory.
       One possible signal of regularization difficulties in the quantum  theory could come from the classical self-energy of a charged particle, which may exhibit more singular behavior  in the noncommutative gauge theory  than appears in  the corresponding commutative theory.   We are, of course, referring to the infinite self-energy of the Coulomb solution.  Thus there is  motivation for studying  properties of the noncommutative analogues of the Coulomb  solution.
    The behavior of the fields and current density of the noncommutative solutions at the origin is of particular interest.  As position eigenstates do not occur in theories with space-space
noncommutativity, there can be no intrinsic notion of  {\it  points} for such theories. It has then been  argued that the point
charges of commutative gauge theories become smeared in noncommutative gauge
theories \cite{Smailagic:2003rp}, but this has not been demonstrated explicitly.  It is noteworthy that such a  desired behavior is seen in Born-Infeld electrodynamics \cite{bi}.  The latter deformation of Maxwell theory has particle-like electrostatic solutions (or bions) which are characterized by a finite size.   Bions are characterized by a finite self-energy and a smooth charge density at the origin (although a singularity in the electric field does remain at the origin).   The Born-Infeld Lagrangian can be given explicitly and a simple expression results for the static solution.   Unfortunately, this is not true for the case of noncommutative electrodynamics, or more precisely, for its equivalent description in terms of commutative gauge fields.   Here we are referring  to the known equivalence of commutative and noncommutative gauge theories expressed through the Seiberg-Witten map \cite{Seiberg:1999vs}.  The Seiberg-Witten map has been given order-by-order in an expansion in the noncommutativity parameter $\theta$.  This then leads to an order-by-order expansion for the effective  Lagrangian expressed in terms of commutative fields.   The zeroth order term (in the case of Abelian gauge fields) is the Maxwell action.   Consequently, at best, one can obtain an   order-by-order
expansion for the field equations and any  particle-like solutions.  The  expansion  in $\theta$  corresponds to an expansion in one over the radial coordinate $r$  for these solutions, and  a  Coulomb-like behavior is expected as $r\rightarrow \infty$.  By going to higher orders in $\theta$ one can probe the intermediate region of the  solutions.  Although the nature of the solutions at the origin remains uncertain in this approach, the behavior of the fields along with the associated self-energy and charge density in intermediate region may provide  clues to whether or not  a singularity is present at  $r=0$. 
   
In this article we shall construct static   solutions to the classical field equations for noncommutative $U(1)$ gauge theory  for three cases: 

$\qquad$ a) $\;$ space-time noncommutativity, 

 $\qquad$ b)  $\;$ space-space noncommutativity and 

$\qquad$  c)  $\;$ both a) and b).

\noindent  The solutions will tend to the Coulomb solution for $r\rightarrow \infty$ and  are valid for intermediate values of $r$.  Constant noncommutativity $\theta$ is assumed for all three cases.  The operator algebra is
  realized on  Minkowski space-time using the standard Gronewald-Moyal star product \cite{groe},\cite{moy}. After applying the inverse Seiberg-Witten map \cite{Seiberg:1999vs}, we obtain the  commutative gauge fields associated with the solutions.   We analyze the self-charge and energy distributions of these solutions for intermediate values of  $r$, and compare the results with Born-Infeld, as well as, Maxwell electrodynamics.
In order to see deviations from the  self-charge and energy distributions of the Coulomb solution it is necessary to carry out the  expansion of the Seiberg-Witten map up to second order in $\theta$.\footnote{ It was shown in \cite{Gaete:2003dh} using a different approach
 that the Coulomb nature of the potential is preserved in noncommutative electrodynamics if one only examines first order effects.}     
The results appear to confirm the above speculation that the classical self-energy singularity in noncommutative electrodynamics is more severe than that of the commutative theory.   For  case b), results from the intermediate region also
appear to indicate that the charge density is better behaved at the origin  than it is in the commutative theory.  On the other hand, the opposite appears  to be true for case a). 
 
 We shall also examine  the effect on the hydrogen atom spectrum of replacing the Coulomb potential  by a space-time noncommutative solution.  Here effects are seen at first order in $\theta$.  (The first order  effects on the spectrum of the space-space noncommutative solution were examined previously in \cite{Stern:2007an},  and  were was used to put experimental bounds on the
noncommutative scale.)   We shall treat the electron in the standard fashion, i.e., applying the standard (commutative) Schr\"odinger equation.  (It was shown in \cite{Balachandran:2004rq},
\cite{Balachandran:2004cr}
that  quantum mechanical spectra are unaffected by replacing the commutative Schr\"odinger equation with one associated with noncommuting time and space coordinates, provided that the spatial coordinates commute.)   Here we can put limits on the space-time noncommutativity parameter.   The bound is a significant improvement over previous results.  We find that space-space and space-time noncommutative sources have distinct experimental signatures.

The outline of this article is the following:
In section 2 we review the Seiberg-Witten map and give expressions up to second order in $\theta$.  We apply it in section 3 to obtain the second order corrections to commutative $U(1)$ gauge theory Lagrangian and Hamiltonian densities, while the static solutions are given in section 4.  In section 5 we  remark on a possible exact expression for the electrostatic Lagrangian for the case  a) of space-time noncommutativity.  If the Lagrangian holds to all orders in the noncommutativity parameter, then we can show that the electrostatic fields of the particle-like solution must be singular at the origin.   In section 6 we apply the lowest order results to the hydrogen atom spectrum.  Concluding remarks are made in section 7.
 We review the Born-Infeld solution in appendix A, while we give  expressions for the nonlinear field equations   and the energy density of noncommutative electrodynamics  in appendix B.

\section{Seiberg-Witten map}
\setcounter{equation}{0}

As it will be essential for the analysis that follows, we here  review the Seiberg-Witten map to second order in the noncommutativity parameter.  We  specialize to
$U(1)$ gauge theory, and denote the commutative
potentials by $a_\mu$, with  gauge variations  $\delta a_\mu
=\partial_\mu\lambda $,    and field strengths $f_{\mu\nu} = \partial_\mu
a_\nu -  \partial_\nu
a_\mu$.  Seiberg  and Witten \cite{Seiberg:1999vs} showed it can be mapped
 to
noncommutative  $U(1)$ gauge theory expressed in terms of noncommutative potentials $A_\mu$
 \be (a,\lambda)\rightarrow
\Bigl(A=A(a),\Lambda=\Lambda(\lambda,a)\Bigr)\label{swmap} \;,\ee
with  gauge variations  \be\delta A_\mu
=\partial_\mu\Lambda -ie[A_\mu, \Lambda]_\star \;,\ee  and field strengths
\be F_{\mu\nu}= \partial_\mu A_\nu - \partial_\nu A_\mu
-ie[A_\mu, A_\nu]_\star \;,\label{fldstrng}\ee where $e$ is the coupling constant and
$[\;,\;]_\star$ is here defined as the star commutator associated with the
Groenewald-Moyal star product\cite{groe},\cite{moy}
\be \star = \exp\;\biggl\{ \frac {i}2 \theta^{\mu\nu}\overleftarrow{
  \partial_\mu}\;\overrightarrow{ \partial_\nu} \biggr\} \label{gmstr} \ee
 $\theta^{\mu\nu}=-\theta^{\nu\mu}$ are constant matrix elements and
  $\overleftarrow{
  \partial_\mu}$ and $\overrightarrow{ \partial_\mu}$ are left and right
derivatives,
respectively, with respect to  coordinates $x^\mu$ on some manifold $M$.   Thus for two function $f$ and $g$ on $M$,  
$[f,g]_\star\equiv f\star g - g\star f$. 
 The coordinates  $x^\mu$ are  associated with
 constant noncommutativity since
\be [x^\mu,x^\nu]_\star= i
\theta^{\mu\nu} \ee   The map (\ref{swmap}) is required to satisfy
\be A_\mu( a + \partial\lambda)-A_\mu(a)  =\partial_\mu\Lambda(\lambda,a)
-ie[A_\mu(a), \Lambda(\lambda,a)]_\star \;, \ee  for infinitesimal transformation parameters $\Lambda$ and $\lambda$.
Solutions for the potentials, fields and transformation parameters can be obtained in terms of expansions in
$\theta^{\mu\nu}$ (or equivalently $e$)
\beqa
A_\mu(a)&=&A_\mu^{(0)}(a)+eA_\mu^{(1)}(a)+e^2A_\mu^{(2)}(a)+\cdot\cdot\cdot\cr
& &\cr  F_{\mu\nu}(a)&=& F_{\mu\nu}^{(0)}(a)+e F_{\mu\nu}^{(1)}(a)+e^2 F_{\mu\nu}^{(2)}(a)+\cdot\cdot\cdot\cr
& &\cr  \Lambda(\lambda,a)&=&\Lambda^{(0)}(\lambda,a)+e\Lambda^{(1)}(\lambda,a)+e^2\Lambda^{(2)}(\lambda,a)+\cdot\cdot\cdot\;,
\label{xpnswmp}
\eeqa  where the zeroth order
correspond to the commutative theory
\be  A_\mu^{(0)}(a) ={ a}_\mu\qquad F_{\mu\nu}^{(0)}(a)=f_{\mu\nu} \qquad  \Lambda^{(0)}(\lambda,a) ={ \lambda}\label{swmpzero}\ee
Explicit expressions up to the second order in $\theta^{\mu\nu}$ have been found by various authors \cite{Jurco:2001rq},\cite{Goto:2000zj},\cite{Brace:2001fj},\cite{Fidanza:2001qm},\cite{Moller:2004qq},\cite{Trampetic:2007hx}.
Up to homogeneous terms, the first and second order  solutions are given by
\beqa    A_\mu^{(1)}(a)& =&\frac
12\theta^{\rho \sigma}{ a}_\rho
 (f_{\mu\sigma}
 -\partial_\sigma a_\mu)  \cr & &\cr  F_{\mu\nu} ^{(1)}(a)& =&-
 \theta^{\rho \sigma}({ a}_\rho\partial_\sigma f_{\mu\nu} + f_{\mu\rho} f_{\sigma\nu}) \cr & &\cr  \Lambda^{(1)}(\lambda,a)& =& \frac 12 \theta^{\mu \nu}
\partial_\mu \lambda a_\nu\label{swmpone} \eeqa and
\beqa
 A_\mu^{(2)}(a)& =&
\frac
{1}2\theta^{\rho \sigma}\theta^{\eta \xi} a_\rho( a_\eta \partial_\xi
f_{\sigma \mu} +\partial_\xi a_\mu\partial_\sigma a_\eta  +
 f_{\sigma \eta}f_{\xi \mu})\cr & &\cr  F_{\mu\nu} ^{(2)}(a)&
 =&\frac{1}2
 \theta^{\rho \sigma}\theta^{\eta \xi}\biggl\{  a_\rho\Bigl(
 \partial_\sigma(a_\eta\partial_\xi f_{\mu\nu}) +
 \partial_\xi f_{\mu\nu} f_{\sigma\eta} + 2 f_{\mu\xi}\partial_\sigma
 f_{\nu\eta}-  2 f_{\nu\xi}\partial_\sigma f_{\mu\eta} \Bigr)+2f_{\mu\sigma} f_{\nu\xi}f_{\rho\eta}\;\biggr\} \cr & &\cr  \Lambda^{(2)}(\lambda,a)& =&  \frac  {1}2\theta^{\rho
  \sigma}\theta^{\eta \xi}\partial_\xi \lambda a_\rho\partial_\sigma a_\eta\label{swmptwo}\;,
\eeqa 
respectively.

\section{Effective Lagrangian}
\setcounter{equation}{0}

Here we give the effective Lagrangian and current density  for noncommutative $U(1)$ gauge theory up to second order in $\theta$.  Although the first order results have been reported previously, the same is not true for the second order results which will be needed for discussions in sec. 4.  The details of the second order analysis appear in Appendix B.  Here we also report on an alternative Lagrangian approach using auxiliary fields. 

 The considerations in the previous section do not involve dynamics. 
In general, (\ref{swmap})  will not map solutions of a commutative gauge theory on a manifold $M$ to
solutions of the corresponding noncommutative gauge
 theory.  Here we choose  $M$ to be four-dimensional Minkowski space-time.  Solutions of the free commutative Maxwell
 equations $\partial^\mu f_{\mu\nu}=0 $  on $M$
 are not in general mapped to   solutions of
 the free noncommutative field equations \be \partial^\mu F_{\mu\nu} - i e[ A^\mu, F_{\mu\nu}]_\star
=0\;,\label{ncfrfldeq} \ee   on $M$.

The converse  is also  true.  Solutions to the free noncommutative field
 equations (\ref{ncfrfldeq}) on $M$ are not
 in general mapped to   solutions to
 the free commutative Maxwell equations on $M$ by the inverse
 Seiberg-Witten map
 \be (A,\Lambda)\rightarrow
\Bigl(a=a(A),\lambda=\lambda(\Lambda,A)\Bigr)\;\label{invswmap} \ee
The field equations (\ref{ncfrfldeq}) are recovered from the
Lagrangian density
\be  L[A] = -\frac 14 F_{\mu\nu}\star F^{\mu\nu} \;,\label{ncglgrn}\ee  which up to boundary terms\footnote{In the next section we will examine solutions with a singularity at the origin.  The boundary terms will affect the singularity.  However,
we shall only be concerned  with the behavior of the fields away from the origin.} is $ -\frac 14 F_{\mu\nu} F^{\mu\nu}$.  From
(\ref{xpnswmp}) it can be expanded in $\theta^{\mu\nu}$ (or equivalently $e$) and expressed in terms of the commutative fields,  giving the effective Lagrangian \be  L[A(a)] = {\cal L}[a] ={\cal L}^{(0)}[a]+e {\cal L}^{(1)}[a] + e^2{\cal L}^{(2)}[a]+\cdot\cdot\cdot\;,\label{fctvlgrn}
\ee    We find that, up to second order, the effective $U(1)$ Lagrangian can be written as a function of only $f_{\mu\nu}$ (and $\theta^{\mu\nu}$),  and not higher derivatives of the fields.  At zeroth order 
we of course recover the  free Maxwell Lagrangian
\be {\cal L}^{(0)}[a] = \frac 14\; {\rm Tr} f^{2}\;,\ee
 while up to total derivatives, the first and second orders are 
given by
\be {\cal L}^{(1)}[a]= -\frac
12 \;{\rm Tr} f^{3}\theta +\frac
1{8}\;{\rm Tr} f^{2}\;{\rm Tr} f\theta\;,\ee
and
\beqa {\cal L}^{(2)}[a] &=& \frac {1}4\; {\rm Tr}( f^{2}\theta)^2+\frac
{1}2 \;{\rm Tr} f^{2} (f\theta)^2 -  \frac {1}4\;{\rm Tr} f^{3}\theta
\;{\rm Tr}f\theta \cr & &\cr & &  +\;\frac {1}{32}\; {\rm Tr}
f^{2}\;({\rm Tr}f\theta)^2 -\frac {1}{16} \; {\rm Tr}
f^{2}\;{\rm Tr}(f\theta)^2 
\eeqa
The first order correction ${\cal L}^{(1)}$ is well known \cite{Gomis:2000sp},\cite{Jurco:2001rq},\cite{Bichl:2001gu}. 
The field  equations resulting from variations of $a$ in (\ref{fctvlgrn}) state that there
is a divergenceless field $B_{\mu\nu}$, 
\be \partial^\mu B_{\mu\nu}=0\;,\label{fctvfeone}\ee
which has and expansion in  $\theta^{\mu\nu}$ (or equivalently $e$) and the field strengths $f_{\mu\nu}$:
\be B_{\mu\nu}= B^{(0)}_{\mu\nu}+ eB^{(1)}_{\mu\nu}+ e^2 B^{(2)}_{\mu\nu}+\cdot\cdot\cdot \;,\label{xpnfrB}\ee
where the zeroth order is just $B^{(0)}_{\mu\nu}=f_{\mu\nu}$.  The first and second orders are computed
in appendix B.  Thus solutions of the noncommutative field equations (\ref{ncfrfldeq}) are mapped under (\ref{invswmap}) to solutions of (\ref{fctvfeone}).

Alternatively, the field equations can be re-expressed in terms of Maxwell equations
for $f$ with an effective conserved current $j_\mu$ associated with the
noncommutative self-interactions
\be \partial^\mu f_{\mu\nu}=j_\nu\;\label{fctvfetwo}\ee
Some properties of these currents have been examined previously in \cite{Banerjee:2003vc},\cite{Banerjee:2005yy}.
After some work we find rather simple expressions for the current in an expansion in $\theta^{\mu\nu}$ (or equivalently $e$) up to order the second order:
\beqa j_\nu&=& e j^{(1)}_\nu+ e^2 j^{(2)}_\nu+\cdot\cdot\cdot \cr & &\cr  j^{(1)}_\nu&=& (f\theta)^{\rho\sigma}\partial_{(\rho}
f_{\sigma)\nu}\cr & &\cr  j^{(2)}_\nu&=&-\Bigl( (f\theta)^2+\frac 12 \theta f^2
\theta \Bigr)^{\rho\sigma}\partial_{(\rho}
f_{\sigma)\nu} \;,\eeqa  
where  parenthesis indicate a symmetrization of indices. 

  The field equations (\ref{fctvfeone}) and (\ref{fctvfetwo}) can also be obtained from the Lagrangian
\be {\cal L}'(B,a) = \frac 12  \;{\rm Tr} Bf - {\cal L}''(B) \;,\label{lgitBa}\ee
where here $B$ as well as $a$ are treated as independent variables and ${\cal L}''(B) $ only depends on $B$.
Up to first order in $\theta^{\mu\nu}$ (or equivalently $e$), the latter is given by
\be  {\cal L}''(B) = \frac 14\; {\rm Tr} B^{2}\; +\;\frac
e2 \;{\rm Tr} B^{3}\theta -\frac
e{8}\;{\rm Tr} B^{2}\;{\rm Tr} B\theta \;+\;{\cal O}(\theta^2) \ee

The Hamiltonian density is obtained from (\ref{fctvlgrn}) in the standard
way
\beqa {\cal H}&=&\frac{\partial {\cal L}}{\partial(\partial_0 a_i)}\;\partial_0 a_i  -
{\cal L} \cr & &\cr &=&{\cal H}^{(0)}+e{\cal H}^{(1)} +e^2{\cal H}^{(2)}+\cdot\cdot\cdot\;,\label{tlnrgdns}
\eeqa where as usual only the spatial components of the potential $a_i$ are dynamical.
The zeroth order term ${\cal H}^{(0)}$ is the Maxwell Hamiltonian, while the first two corrections are computed in  appendix B.

\section{Static solutions}
\setcounter{equation}{0}

Here we look for static perturbative solutions to the  field equations (\ref{fctvfeone}) or (\ref{fctvfetwo}) on four-dimensional Minkowski space $M$.  (More precisely,  $M$ is defined with the spatial origin removed.)   The  perturbative expansion  in $\theta^{\mu\nu}$ also corresponds to an expansion in one over the radial coordinate $r$  for the solutions.  The result can therefore be considered  valid for intermediate values of  $r$. Rather than solve the commutative field equations (\ref{fctvfeone}) or (\ref{fctvfetwo}) directly,  it is simplest to first solve  (\ref{ncfrfldeq}) for the noncommutative potentials $ A_\mu$ and then apply the inverse Seiberg-Witten map (\ref{invswmap}) to get the commutative potentials $a_\mu$.   This is the approach we follow below.

\subsection{Space-time noncommutativity}

Here we consider $\theta^{0i}\ne 0$ and $\theta^{ij}= 0$, where $i$ is a space index and $0$ the time index.  We shall be concerned with electrostatic fields.   
The star commutator vanishes when acting between any static fields in this case, and so  the noncommutative gauge field equations (\ref{ncfrfldeq}) reduce to the commutative Maxwell equations.  The Coulomb solution 
\be A_0  = -\frac er \qquad\quad A_i=0 \label{thzislnone}\;,\ee is then exact in this case.
Upon performing the inverse Seiberg-Witten map (\ref{invswmap}) of this solution  one  gets the following results for the commutative potentials and field strengths
\beqa a_0& =& -\frac er\;-\; \frac{e^3\theta^{0i} \hat x_i}{r^3}\;+\;
\frac{e^5}{2r^5}\Bigl((\theta^{0i})^2 -5 (\theta^{0i}\hat x_i)^2\Bigr)\;+\;\cdots\cr & & \label{stncinvswcoul}\\  f_{i0}&=& \frac {e\hat x_i}{r^2} \biggl( 1 +\frac {4e^2}{r^2}\theta^{0j}\hat x_i - \frac{5e^4}{2r^4}\Bigl( (\theta^{0i})^2 - 7(\theta^{0j}\hat x_j)^2\Bigr)+\cdots\biggr) - \frac {e^3\theta^{0i}}{r^4} \biggl(1 + \frac {5e^2}{r^2}\theta^{0j}\hat x_i +\cdots\biggr)\nonumber\;, 
 \eeqa along with $a_i= f_{ij}=0$.  (The hat denotes a unit vector $\hat x_i=x_i/r$.) 
 For this solution we can identify the   current density in the Maxwell equation (\ref{fctvfetwo}) with
 \be j_0=- \frac { 4 e^3}{r^5} \theta^{0j}\hat x_j \;+\;\frac{5e^5}{r^7}\Bigl( (\theta^{0i})^2 -7 (\theta^{0i}\hat x_i)^2\Bigl) \;+\; {\cal O}(\theta^3)\;, \quad\qquad j_i =0\;, \qquad r>0\;, \label{crntdnstst} \ee  which  breaks rotational invariance.
 From Gauss' law the resulting effective charge inside a sphere of radius $r$ centered about the origin is
 \be \frac 1 {4\pi}\int d\Omega \;r^2 \hat x_i f_{i0} =  e \Bigl( 1\;+\; \frac 53\frac{ e^4}{ r^4} (\theta^{0i})^2\;+\;{\cal O}(\theta^3) \Bigr)\;,\label{chrgcrst}\ee which increases with decreasing $r$.  $\Omega$ is the solid angle. This is in contrast to what happens for the electrostatic solution (bion) of Born-Infeld theory [see (\ref{bichrgdst})].  The bion  charge inside a sphere of radius $r$ goes smoothly to zero as $r\rightarrow 0$.  
In contrast, a singular source  may be a general feature of gauge theories with  space-time noncommutativity. Moreover, the singularity appears to be more severe than that which occurs in the commutative theory.   The result (\ref{chrgcrst}) may indicate that the exact solution  screens an infinite charge at the origin.  

  For the case of  $\theta^{ij}=0$ and electrostatic fields, the Lagrangian density (\ref{fctvlgrn}) and energy density (\ref{tlnrgdns}), simplify  to 
 \beqa {\cal L}_{es}^{\theta^{ij}=0}&=&  \frac 12 (f_{i0})^2\Bigl( 1 -  e(\theta^{0j}f_{j0}) +  {e^2}(\theta^{0j}f_{j0})^2+\cdots \Bigr)\label{elstclgn}\\ & &\cr{\cal H}_{es}^{\theta^{ij}=0}&=&  \frac 12 (f_{i0})^2\Bigl( 1 -  2e(\theta^{0j}f_{j0}) +3  {e^2}(\theta^{0j}f_{j0})^2+\cdots \Bigr) \;,\label{elstcham}
 \eeqa respectively.
  Substituting  the solution (\ref{stncinvswcoul}) in the energy density (\ref{elstcham}) and averaging over a sphere of radius $r$ gives
 \be  \frac 1 {4\pi}\int d\Omega \;{\cal H} = \frac {e^2}{2r^4}\biggl( 1 \;+\;  \frac{10}3  \frac {e^4}{r^4} (\theta^{0i})^2\;+\; {\cal O}(\theta^3)\biggr) \ee
This is also in contrast to what happens for the bion [see (\ref{bionnrg})], which has a finite self-energy.  For the noncommutative solution (\ref{stncinvswcoul}), the self-energy in the intermediate region  grows faster than that for a Coulomb point charge  for decreasing $r$.

 An alternative approach to obtaining  solution (\ref{stncinvswcoul}) would be to solve the  field equations (\ref{fctvfeone}) directly.  Since we are interested in electrostatic field configurations on ${\mathbb{R}}^{3}$ minus the origin we may try setting the  divergenceless field $B_{\mu\nu}$ equal  to the Coulomb solution.  More generally, one can (actually, must) add homogeneous terms, in analogy to a multi-moment expansion:
 \be B_{i0} = \frac {e\hat x_i}{r^2}   + c_1 \frac {e^3}{r^4}\Bigl( \theta^{0i} - 2  \theta^{0j}\hat x_j\hat x_i\Bigr)  + c_2 \frac {e^5}{r^6}\Bigl((\theta^{0j})^2 \hat x_i  + 4 \theta^{0j}\hat x_j\theta^{0i} -7 (\theta^{0j}\hat x_j)^2\hat x_i  \Bigr)+\cdots \;,\label{homplsin}\ee 
 where $c_a$ are arbitrary coefficients and $ B_{ij}=0$.  Starting from the Lagrangian density (\ref{elstclgn}) for electrostatic fields one obtains the following expansion up to second order for the divergenceless fields $B_{i0}$ in terms of $f_{i0}$ 
 \be B_{i0} = f_{i0}\Bigl( 1 -  e(\theta^{0j}f_{j0}) +  {e^2}(\theta^{0j}f_{j0})^2+\cdots \Bigr) - e \theta^{0i}(f_{k0})^2\Bigl( \frac12 -  e(\theta^{0j}f_{j0}) +\cdots \Bigr) \ee  
 We can  invert this expression to solve for the field strengths $f_{i0}$ for the general solution (\ref{homplsin}).  The result is 
\beqa f_{i0}& = &\frac {e\hat x_i}{r^2}   + \Bigl(c_1 + \frac 12\Bigr)\frac {e^3}{r^4} \theta^{0i}  +(1-2 c_1) \frac {e^3}{r^4}\theta^{0j}\hat x_j\hat x_i + \Bigl(c_1+c_2+\frac 12\Bigr) \frac {e^5}{r^6}(\theta^{0j})^2 \hat x_i \cr & &\cr & &+ (4c_2+1) \frac {e^5}{r^6} \theta^{0j}\hat x_j\theta^{0i}+ (1-4c_1 -7c_2) \frac {e^5}{r^6} (\theta^{0j}\hat x_j)^2\hat x_i  +\cdots  \;,\eeqa along with  $ f_{ij}=0$.   Lastly, if we impose the Bianchi identities,
which here means $\partial_if_{j0}= \partial_jf_{i0}$, we obtain the  unique result up to second order that $c_1=c_2= -\frac 32$, and thus recover the solution (\ref{stncinvswcoul}).

\subsection{Space-space noncommutativity}

Here we consider $\theta^{0i}= 0$ and $\theta^{ij}\ne 0$.
 A static solution of the free
noncommutative field equations (\ref{ncfrfldeq})  on  
${\mathbb{R}}^{3}$ (minus a point) $\times$ time was obtained perturbatively in
\cite{Stern:2007an}. It is   given by
\be A_0  = -\frac er\;+\;\frac{e^5}{4r^5}\;\biggl\{-(\theta^{ki} \hat x_i)^2+\frac 15 (\theta^{ij})^2  \biggr\} 
\;+\;{\cal O}(\theta^3)\qquad\quad  A_i =\frac{e^3\theta^{ij}\hat x_j}{4r^3}\;+\;{\cal O}(\theta^3) \label{nccoul}\ee Upon performing the inverse Seiberg-Witten map (\ref{invswmap}) of this solution to the commutative potentials and field strengths one  gets
\beqa a_0(A) = -\frac er\;+\;
\frac{e^5(\theta^{ij})^2}{20r^5}  \;+\;{\cal O}(\theta^3) &\qquad  & a_i(A) =\frac{e^3\theta^{ij}\hat x_j}{4r^3}\;+\;{\cal O}(\theta^3)\;,\cr & & \label{ssncinvswcoul}\\
 f_{i0} = \frac {e\hat x_i}{r^2} \;  -\;\frac{e^5(\theta^{ij})^2\hat x_i}{4r^6} \;+\;{\cal O}(\theta^3) &\qquad & f_{ij}= \frac{e^3}{r^4} \Bigl(-\frac 12 \theta^{ij} +\theta^{ik}\hat x_k\hat x_j -\theta^{jk}\hat x_k\hat x_i\Bigr)\;+\;{\cal O}(\theta^3)\nonumber \eeqa
Unlike in the previous case, this is not an electrostatic solution of the field equations (\ref{fctvfetwo}) as magnetic fields are present (both in the commutative and noncommutative theory).  Up to order $\theta^2$, the charge density  is  rotationally invariant, while the current density is not
 \be j_0= \frac {  e^5}{r^7} (\theta^{ij})^2 \;+\; {\cal O}(\theta^3)\quad\qquad j_i =  \frac {  e^3}{r^5} \theta^{ij}\hat x_j\;+\; {\cal O}(\theta^3)\;,\qquad r>0\label{crntdnstss}
 \ee 
The resulting effective charge inside a sphere of radius $r$ centered about the origin is 
 \be \frac 1 {4\pi}\int d\Omega \;r^2 \hat x_i f_{i0} =  e \Bigl( 1\;-\; \frac{ e^4 (\theta^{ij})^2}{4 r^4}\;+\;{\cal O}(\theta^3)\Bigr)\;,\label{chrgcrss}\ee which here decreases with decreasing $r$.  This behavior is similar to that of the bion (\ref{bichrgdst}).  It is then a possibility that the  charge distribution for the theory with space-space noncommutativity is smooth at the origin, and that one can associate the size of the distribution with
$e\sqrt{ |\theta^{ij}|}$, as has been done previously \cite{Smailagic:2003rp}.

 Substituting into the energy density computed in the appendix (\ref{Hzrontw}) and averaging over a sphere of radius $r$ now gives
 \be  \frac 1 {4\pi}\int d\Omega \;{\cal H} = \frac {e^2}{2r^4}\biggl( 1 \;+\;  \frac{13}{24}  \frac {e^4}{r^4} (\theta^{ij})^2\;+\; {\cal O}(\theta^3)\biggr)\; \ee
 Once again, in contrast to the energy density of the Born-Infeld solution (\ref{bionnrg}), it grows faster than the self-energy of a Coulomb point charge  as $r$ tends to the origin.

\subsection{General case}

The solution (\ref{nccoul}) to the free
noncommutative field equations (\ref{ncfrfldeq}) applies in the general case where both $\theta^{0i}$ and $\theta^{ij}$ are nonzero.
Upon performing the inverse Seiberg-Witten map (\ref{invswmap}) of (\ref{nccoul}) in this case, one  gets
\beqa a_0(A)& =& -\frac er\;-\; \frac{e^3\theta^{0i} \hat x_i}{r^3}\;+\;
\frac{e^5}{2r^5}\Bigl( \frac 1{10} (\theta^{ij})^2+(\theta^{0i})^2 -5 (\theta^{0i}\hat x_i)^2\Bigr)\;+\;{\cal O}(\theta^3) \cr & &\cr  a_i(A)& =&\quad\frac{e^3\theta^{ij}\hat x_j}{4r^3}\;+\;
\frac{e^5}{8r^5}\theta^{0j}\Bigl(3 \theta^{ji} - 3\theta^{jk}\hat x_i\hat
x_k + 8 \theta^{ik} \hat x_j\hat x_k\Bigr)\;+\;{\cal O}(\theta^3)\; \label{invswcoul}\eeqa
The charge density for the general case is simply the sum of  (\ref{crntdnstst}) and (\ref{crntdnstss}) up to second order, while the current density  contains mixed terms at second order; i.e. terms proportional to the product of both $\theta^{ij}$ and $\theta^{0k}$.  
To obtain the effective charge inside a sphere of radius $r$ up to second order in  the general case, one adds the contributions (\ref{chrgcrst}) and (\ref{chrgcrss}).  These second order contributions cancel when $(\theta^{0i})^2 = \frac 3{20}(\theta^{ij})^2$.  Further analysis of the general case is quite involved and does not appear to lead to novel results.

\section{Exact electrostatic Lagrangian for $\theta^{0i}\ne0$, $\theta^{ij}=0$?}

This section is  speculative in nature and  offers a possible nonperturbative treatment for case a), and if correct, gives the   solution for all $r$.  Having an exact Lagrangian could also lead to an investigation of solutions which do not possess a commutative limit.

From the second order  results  for the electrostatic Lagrangian (\ref{elstclgn}) and energy density (\ref{elstcham}) for the case  $\theta^{0i}\ne0$, $\theta^{ij}=0$, it is tempting to surmise that the exact expressions for the electrostatic Lagrangian  and energy density in this case are 
\be {\cal L}_{es}^{\theta^{ij}=0}= \frac{ \frac 12 (f_{i0})^2}{ 1 +  e\theta^{0j}f_{j0}} \label{xctesstL}\ee
and 
\be {\cal H}_{es}^{\theta^{ij}=0}= \frac{ \frac 12 (f_{i0})^2}{( 1 +  e\theta^{0j}f_{j0})^2}\;\;, \ee
respectively.  The field equation following from (\ref{xctesstL}) states that 
\be B_{i0} =\frac{f_{i0}}{ 1 +  e\theta^{0j}f_{j0}} - \frac{\frac e2 \;\theta^{0i}(f_{k0})^2}{( 1 +  e\theta^{0j}f_{j0})^2} \label{Bfrxt}\label{esxrltnon}\ee has zero divergence (\ref{fctvfeone}).  This can be inverted  to obtain an expression for  the field strength
\be f_{i0} = \frac {B_{i0}\; +\; \frac {\theta^{0i}}{e(\theta^{0n})^2}\Bigl(1-e\theta^{0m}B_{m0}\Bigr)}{\sqrt{\Bigl(1-e\theta^{0j}B_{j0}\Bigr)^2 -\;e^2 (\theta^{0\ell})^2(B_{k0})^2}} \;\;-\;\; \frac {\theta^{0i}}{e(\theta^{0\ell})^2}\label{esxrltnton}\ee
 The field equations  can then also be obtained from the Lagrangian
\be {\cal L}_{es}^{'\theta^{ij}=0}(B,a) =\frac 1{e^2 (\theta^{0n})^2} {\sqrt{\Bigl(1-e\theta^{0j}B_{j0}\Bigr)^2 -\;e^2 (\theta^{0\ell})^2(B_{k0})^2}}\;+\; \frac {\theta^{0i}B_{i0}}{e (\theta^{0n})^2}\; +\; B_{i0} f_{i0}\;,\ee
where $B$ and $a$ are treated as independent variables as in (\ref{lgitBa}).
From (\ref{esxrltnton}) a well defined solution should everywhere satisfy
\be  e |\theta^{0\ell}| |B_{k0}|+e\theta^{0j}B_{j0}\;\le\; 1 \label{cntcnd}\ee
So, for example, we cannot simply set the divergenceless field $B_{i0}$ proportional to the Coulomb term $\hat x^i/r^2$.  The solution must include homogeneous terms as in (\ref{homplsin}).  This is also needed for $f_{i0}$ to satisfy  the Bianchi identity. 

Although we have not found a nontrivial  exact solution to the electrostatic field equation $\partial_i B_{i0}=0$ which goes to the Coulomb solution as $r\rightarrow \infty$, we can show that that there are no regular solutions near the origin.   The proof is by contradiction.  Assume a power law behavior  for the electrostatic potential near $r=0$  
\be a_0 \sim \gamma\;\frac{ r (\theta^{0i}\hat x_i)}{e (\theta^{0k})^2}  +\beta  r^n (\theta^{0i}\hat x_i)^m \;,\label{asmpnrz}\qquad {\rm as}\;r\rightarrow 0\;, \ee where $\gamma $ and $\beta $ are constants and $n>1$ and $m\ge 0$.  Then to leading order in $r$, the field strength tends to a constant 
\be f_{i0} \sim  \frac{\gamma  \theta^{0i}}{e (\theta^{0k})^2}\;  +\; \beta \;r^{n-1} \Bigl[(n-m) (\theta^{0j}\hat x_j)^m \hat x_i + m (\theta^{0j}\hat x_j)^{m-1}\theta^{0i} \Bigr]\;,\qquad {\rm as}\;r\rightarrow 0\ee 
Upon substituting into (\ref{Bfrxt}) and computing the divergence to leading order in $r$, we get
$$ \partial_i B_{i0}\; =\;\frac{\beta r^{n-2}(\theta^{0i}\hat x_i)^{m-2}} {1+\gamma } \Biggl\{ { (n-m) (n+m+1) (\theta^{0j}\hat x_j)^2 + m(m-1)  (\theta^{0j})^2}  $$ \be -\; \frac{ (2 +\gamma )\gamma }{(1+\gamma )^2(\theta^{0k})^2}\Bigl((n-m)(n-m-2)(\theta^{0j}\hat x_j)^4 +(n-m)(1+2m)(\theta^{0j}\hat x_j)^2(\theta^{0\ell})^2 + m(m-1) (\theta^{0\ell})^4 \Bigr)\Biggr\} \ee
A regular asymptotic  solution requires that all the coefficients of $(\theta^{0j}\hat x_j)^N$  in the braces vanish.  From the vanishing of the $N=4$ coefficient one gets a) $n=m$ or b) $n=m+2$.   From the vanishing of the $N=0$ coefficient one gets $m=0$ or $1$, which for a) is inconsistent with $n>1$, and for b) leads to $\gamma$ not real (after demanding  that the $N=2$ coefficient vanishes).  There are then no nontrivial asymptotic solutions of the form (\ref{asmpnrz}) near the origin.

\section{Application to the hydrogen atom}

The effect of a space-space noncommutative source on the hydrogen atom  at lowest order in $\theta^{ij}$ was considered  in \cite{Stern:2007an}, and it was used to put experimental bounds on the
noncommutative scale.  We first  briefly review the argument below and then make analogous arguments for the case of space-time noncommutativity.  We obtain a bound for the space-time noncommutativity parameter which is a substantial improvement over previous attempts.

  We start with the standard nonrelativistic Hamiltonian for a charged spinning particle
\be H =  \frac 1{2m_e} ( p_i- e a_i)^2 +e a_0 -\mu_B \epsilon^{ijk} f_{ij}S_k  \;,\ee 
where $m_e$ is the electron mass,   $\mu_B=\frac e{2m_e}$ denotes the Bohr  magneton and we use natural units  $\hbar=c=1$.  For the case of  space-space noncommutativity at leading order, $a_0$  is the Coulomb potential and  $a_i$ and $f_{ij}$ are the first order expressions given in (\ref{ssncinvswcoul}).  Then \be  H= H_0+H^{(ss)}_1+H^{(ss)}_2 \;,\ee where $H_0$ is the nonrelativistic hydrogen atom Hamiltonian \be H_0= \frac 1{2m_e}  p_i^2 -\frac {\alpha}r\;,\ee and $H^{(ss)}_1$ and $H^{(ss)}_2 $
are the leading perturbations
\be H^{(ss)}_1= \frac{\alpha^2}{4 m_e} \frac {\vec \theta\cdot \vec L}{r^4}\;, \qquad H^{(ss)}_2= \frac {\alpha^2}{2m_e}\; \frac{2(\vec S\cdot\hat x)(\vec \theta\cdot\hat x)-\vec S\cdot \vec \theta }{r^4}\ee  where $\theta^{ij} = \epsilon^{ijk} \theta _k$ and $\alpha$ is the fine structure constant.  $H^{(ss)}_1$ leads to corrections to the Lamb shifts of the $\ell\ne 0$ states, while $H^{(ss)}_2 $ induces splittings in the $1s$ states.  After choosing $\vec \theta$ in the third direction  $\theta_{i} = \theta_{ss}\delta_{i3}$, one gets the following diagonal matrix elements
\beqa 
 <H^{(ss)}_1>_{2P_{1/2}^{\pm 1/2}}&=&\frac{\alpha^2\theta_{ss} }{4m_e}\;\biggl<\frac {  L_z}{ r^4 }\biggr>_{2P_{1/2}^{\pm 1/2}}=\;\pm\;\frac{\alpha^6m^3_e\theta_{ss}}{144}\;\label{splttwo}\\
 & &\cr <H^{(ss)}_2 >_{1S_{1/2}^{\pm 1/2}}&=& -\;\frac{\alpha^2 \theta_{ss}}{6m_e}\;\biggl<\frac{S_z}{r^4 }\biggr>_{1S_{1/2}^{\pm
 1/2}}\;=\;\mp\frac 13{\alpha^5m_e^2\theta_{ss}}{\Lambda_{QCD}}\label{ncrtnsthpf}
\;,\eeqa 
using spectroscopic notation $n\ell_{j}^{m_j}$.   To get a finite answer for (\ref{ncrtnsthpf}) we  inserted the $\Lambda_{QCD}$
 cutoff, taking into account the finite size of the
 nucleus.  (This appears to be the best one can do  without having a consistent
  treatment of noncommutative QCD.) According to
  \cite{Eides:2000xc} the current theoretical accuracy on the $2P$ Lamb
  shift is about $0.08$ kHz.   From the splitting (\ref{splttwo}),
  this then gives the  bound
\be \theta_{ss}\; {}^<_\sim\;(30\; {\rm  MeV})^{-2} \ee   The current theoretical accuracy on the $1S$ 
  shift is about $14$ kHz \cite{Eides:2000xc}.  From the splitting (\ref{ncrtnsthpf}) and $\Lambda_{QCD}\sim 200$ MeV,
 this then gives the improved bound of \be \theta_{ss}\; {}^<_\sim \; (4\; {\rm GeV})^{-2} \ee 
 
 In the above treatment of  the hydrogen atom we considered the gauge fields generated by the proton to be noncommutative, but we applied it to the standard Schr\"odinger equation, rather than its noncommutative counterpart.  The relative coordinates  of the Schr\"odinger equation were treated as commuting.  In this regard, it has been noted that although the noncommutativity parameter is fixed in quantum field theory, this is not necessarily the case in noncommutative quantum mechanics where  the different particle coordinates may be associated with different $\theta^{\mu\nu}$ \cite{Ho:2001aa},\cite{Chaichian:2000si}.   For a multi-particle quantum system, the commutation relations for the different particles should, in
principle, be derived starting from the noncommutative field
theory. 
 In \cite{Ho:2001aa} starting from a noncommutative version of
QED the authors found that two Dirac particles of opposite charge have
opposite noncommutativity, while the relative coordinates commute.
As pointed out in \cite{Chaichian:2000si}, the correct approach for the hydrogen atom would
have to include noncommutative  QCD, which unfortunately is not well
understood.   A  pragmatic
approach would be to instead  set separate bounds on the
noncommutativity of the electron and nucleus. 

 In contrast to the above approach, the spatial coordinates do not commute in the space-space noncommutative version of the Schr\"odinger equation.  The latter was considered previously for the hydrogen atom in \cite{Chaichian:2000si}, and it led to the additional perturbation
 \be  H^{(e)}=\;-\;\frac {\alpha }{2 }\;\frac { \vec \theta \cdot
  \vec L}{ r^3 }\ee to $H_0$ and  a further correction to the $2P$ Lamb
  shift.   This gives yet another bound on the noncommutativity parameter $\theta_{i} = \theta_{e}\delta_{i3}$, here associated with the electron.  The bound was correctly computed in \cite{Stern:2007an} to be 
  \be  \; \theta_e\; {}^<_\sim \; (6\; {\rm GeV})^{-2}\; \ee

 It remains  to find the effect of a space-time noncommutative source on the hydrogen atom.  Here we replace the standard Coulomb potential for the hydrogen atom with $ea_0$, with $a_0$ given by  (\ref{stncinvswcoul}).  We shall again only be interested in leading order effects.  
The hydrogen atom Hamiltonian is then
\be  H= H_0+H^{(st)}\;,\qquad H^{(st)}=-\; \frac{\alpha^2\theta^{0i} \hat x_i}{r^3}\; \ee  
Unlike with the case of space-space noncommutativity, there are no diagonal matrix elements amongst the orbital angular momentum eigenstates for the perturbative term $H^{(st)}$. Choosing $\theta^{0i} = \theta_{st}\delta_{i3}$, the latter gives rise to nonvanishing matrix elements between states with $\Delta \ell=1$ and $\Delta m=0$.  The matrix element between the degenerate $2$s and $2$p  $m=0$ states is
\be <n=2,\ell=0,m_L=0 |\;H^{(st)}\;|n=2,\ell=1,m_L=0>\; =\; -\frac{\alpha^5 m_e^3\theta_{st}}{27} \;\ee  Upon including  the electron spin (and the fine structure), this implies a mixing of the four degenerate $n=2, j=\frac 12$ states (i.e., $2s_{1/2}$ and $2p_{1/2}$)
\be <2s_{1/2}^{\pm 1/2}|\;H^{(st)}\;|2p_{1/2}^{\pm 1/2}>\; =\; \pm \frac{\alpha^5 m_e^3\theta_{st}} {27\sqrt{3}} \;,\ee   Consequently, there is an energy splitting of the levels equal to 
\be \Delta E=\frac{2\alpha^5 m_e^3\theta_{st}}{27\sqrt{3}} \label{destnc} \;,\ee and two sets of doubly-degenerate levels result.  Then from the above mentioned  current theoretical accuracy on the $2P$ Lamb
  shift and (\ref{destnc}), one gets the bound \be \theta_{st}\; {}^<_\sim \; (.6\; {\rm GeV})^{-2}\label{bndnstnc}  \ee
  
  Previous bounds on the  space-time noncommutativity parameter have been found using gravitational quantum well experiments \cite{OBJGRCMLDZ},\cite{OBJGR},\cite{RBBDRSS},\cite{Saha:2006xx}.  (\ref{bndnstnc}) is a significant improvement.
Unlike with the case of space-space noncommutativity at lowest order, all of the $n=2, j=\frac 12$ hydrogen  atom states  are shifted.  Thus the experimental signature for space-time noncommutativity differs from that for space-space noncommutativity.
 
 As before we considered the gauge fields generated by the proton to be noncommutative, but  we applied the standard (commutative) Schr\"odinger equation for the electron.  $x_i$ does  not commute with $t$ in the space-time noncommutative version of the Schr\"odinger equation.  However, it was shown in \cite{Balachandran:2004rq},\cite{Balachandran:2004cr}
that the   quantum mechanical spectrum  is unaffected by the replacement of the commutative Schr\"odinger equation with the one associated with noncommuting time and space coordinates (provided that the spatial coordinates commute).

\section{Conclusions}

We have constructed perturbative static solutions to the classical field equations of noncommutative $U(1)$ gauge theory up to second order in $\theta$ for the three cases: a)  space-time noncommutativity, b) space-space noncommutativity c) both a) and b).  They tend to the Coulomb solution as $r\rightarrow \infty$.  For case a) the solution is electrostatic and the associated self-charge inside a sphere of radius $r$ centered about the origin increases with decreasing $r$.  This may signal that the exact solution screens an infinite charge at the origin.  We proposed an exact expression  for the Lagrangian  and Hamiltonian in this case and, if it is correct, have shown that no nonsingular solutions exist at the origin.  Magnetic as well as electric fields are present for the case b) solution, and here the self-charge inside a sphere of radius $r$ centered about the origin  decreases with decreasing $r$.  It then becomes plausible that the charge density is well behaved at the origin, as happens in Born-Infeld electrodynamics.  If so, the guess  that
charges become smeared in  gauge
theories with space-space noncommutativity would be valid.
  For both cases a) and b)  the self-energy of the solutions in the intermediate region  grows faster than that for a Coulomb point charge as $r$ tends to the origin.  It thus appears that the noncommutative solutions have infinite self-energy, contrary to the case of Born-Infeld solution, and that the divergence of the classical self-energy in the noncommutative theory is more severe than its counterpart in the commutative theory.

We have also looked for the lowest order effects of these solutions on the hydrogen atom spectrum  and used them to put experimental bounds on the space-time and space-space
noncommutative scales.  We found that the two different cases have different experimental signatures.

It is known that the star product realization of any given operator algebra on a noncommutative space is not unique.  For example, for the case of constant noncommutativity one can also apply the Voros star product which is based on coherent states \cite{vor},\cite{Alexanian:2000uz}.
   More generally,  the star product  
belongs to a very large equivalence class  of star
products \cite{Kontsevich:1997vb}.  The different star     products
in  the equivalence class are related by gauge transformations,  and tools have been developed for writing down gauge theories based on these equivalence classes \cite{Pinzul:2007bk}.  Since the analysis in this article relies on one particular choice of the star product; i.e., the Groenewald-Moyal star product, it is important to know how sensitive the above results are to this choice.  It would be especially puzzling if the results obtained for the hydrogen atom spectrum depended on the choice of  star product.

\bigskip

\noindent
{\bf Acknowledgment}

\noindent
I am grateful to  Aleksandr Pinzul
for useful discussions.

\bigskip

\appendice{{\bf Comparison with the bion}}

 It is useful to compare the asymptotic behavior of  the charge density and self-energy of the solutions found in section 4 with that of a known deformation of Maxwell theory, i.e., Born-Infeld theory. The Born-Infeld Lagrangian
\cite{bi}  on four-dimensional Minkowski space is constructed from the determinant of the
matrix 
\be h =\eta + \kappa f\;,\ee
 where $f=[f_{\mu\nu}]$ is the field tensor, 
 $\eta=[\eta_{\mu\nu}]={\rm diag}(-1,1,1,1)$ is the flat metric tensor and $\kappa$ is a dimensionful constant. (In string theory the latter is identified with $2\pi$ times the string constant.)  The Lagrangian density is 
\beqa  {\cal L}^{BI} &=&{\kappa}^{-2} \biggl(1 -
\sqrt{-\det h} \biggr) \label{tbia} \;, \cr & &\cr \det h&=& -1 - \frac{\kappa^2}2 f_{\mu\nu}f^{\mu\nu} + \frac{\kappa^4}{64}(\epsilon^{\mu\nu\rho\sigma }f_{\mu\nu}f_{\rho\sigma})^2\label{bilgrngn}
\eeqa
One recovers the Maxwell action at lowest order in the  expansion in $\kappa$. 
The  field  equations resulting from (\ref{bilgrngn}) state that there
is a divergenceless field $B^{BI}_{\mu\nu}$, analogous to $B_{\mu\nu}$ in (\ref{fctvfeone}),
\be \partial^\mu B^{BI}_{\mu\nu}=0\;,\qquad B^{BI}_{\mu\nu}= \frac{1}{\sqrt{-\det h}}\;\Bigl( f_{\mu\nu} - \frac{\kappa^2}{16} (\epsilon^{\alpha\beta\gamma\delta}f_{\alpha\beta}f_{\gamma\delta})(\epsilon_{\mu\nu\rho\sigma }f^{\rho\sigma})\Bigr)\label{sstcflde}\ee
Alternatively, the field equations  can be re-cast as Maxwell equations
for the field strength $f_{\mu\nu}$ with an effective conserved current $j^{BI}_\nu$ as in (\ref{fctvfetwo})
\be \partial^\mu f_{\mu\nu}=j^{BI}_\nu\;,\label{bime}\ee where here $j^{BI}_\nu=\partial^\mu (f_{\mu\nu}-B^{BI}_{\mu\nu})$. 
The energy density is given by \be {\cal H}^{BI}= \frac{1}{\sqrt{-\det h}}\;\Bigl( \frac{1}{2}(f_{ij})^2+\frac{1}{\kappa^2}\Bigr) \;-\;\frac{1}{\kappa^2}\label{bihamden}\;,\ee  and it is easy to check that the Maxwell energy density is recovered at zeroth order in a  
$\kappa$ expansion.

For the case of electrostatics, (\ref{tbia}) and (\ref{bihamden}) reduce to 
\beqa  {\cal L}^{BI}_{es} &=&{\kappa}^{-2} \biggl(1 -
\sqrt{1-\kappa^2(f_{i0})^2} \biggr)  \cr & \cr{\cal H}^{BI}_{es} &=&{\kappa}^{-2} \biggl(\frac 1{
\sqrt{1-\kappa^2(f_{i0})^2}}-1 \biggr)\;,\eeqa respectively. Substituting the Coulomb solution for  $B^{BI}_{\mu\nu}$ on  
${\mathbb{R}}^{3}$ (minus a point $)\;\times$ time
\be B^{BI}_{i0}=  e \frac{\hat x_i}{r^2}\qquad  B^{BI}_{ij}=0\;,\ee
gives the bion solution for $f_{\mu\nu}$ \be  f_{i0} = \frac{e\hat x_i}{\sqrt{r^4 + \kappa^2 e^2 }} \qquad  f_{ij}=0\;, \ee 
which satisfies the Bianchi identities $\partial_if_{j0}= \partial_jf_{i0}$.  The associated electrostatic potential can be expressed in terms of a hypergeometric function.\cite{Pinzul:2003vp}  Using (\ref{bime}) one then gets the following continuous charge distribution for the solution
\be j_0^{BI} =\frac{2\kappa^2 e^3}{r(r^4 + \kappa^2 e^2 )^{3/2}} \ee   
and the following effective charge inside a sphere of radius $r$ centered about the origin 
 \beqa \frac 1 {4\pi}\int d\Omega \;r^2 \hat x_i f_{i0} &=& \frac{e}{\sqrt{1 + \bigl(\frac{\kappa e}{r^2}\bigr)^2}}\rightarrow \left\{\matrix{ e\biggl( 1\;-\;\frac{\kappa^2 e^2}{2r^4}\;+\;\cdots\biggr)& {\rm as} &r\rightarrow\infty\cr  \frac{r^2}\kappa \biggl( 1\;-\; \frac{r^4} {2\kappa^2 e^2}\;+\;\cdots\biggr)  & {\rm as} &r\rightarrow 0\cr}\right.\;,\label{bichrgdst}\eeqa which  decreases monotonically to zero as $r$ goes to zero.   Substituting into (\ref{bihamden}) gives the bion energy density \beqa  {\cal H}^{BI}_{es} &=&\frac{1}{\kappa^2}\biggl(\sqrt{1+\Bigl(\frac{\kappa e}{r^2}\Bigr)^2}\;-\;1\biggr)\cr & &\cr
 &\rightarrow &\frac {e^2}{2r^4}\biggl( 1 \;-\;\frac{\kappa^2 e^2}{4r^4} \;+\;\cdots\biggr)\qquad {\rm as}\;\;r\rightarrow\infty \;,\label{bionnrg}\eeqa
whose integral is finite.  
\bigskip

\appendice{{\bf Sourceless field $B_{\mu\nu}$ and Hamiltonian density}}

Here we give explicit  expressions for i) the sourceless field $B_{\mu\nu}$ in the noncommutative field theory, along with ii) the Hamiltonian density in section 3, up to second order in $\theta^{\mu\nu}$.

i) Varying the commutative field strengths $f_{\mu\nu} $ in the Lagrangian density (\ref{fctvlgrn}) gives
\be \delta{\cal L} =\frac 12 \; {\rm Tr} B \delta f \;,\ee
and the resulting field  equations  state that  $B_{\mu\nu}$ satisfies (\ref{fctvfeone}). $B_{\mu\nu}$
is expanded in terms of $\theta^{\mu\nu}$ ( or equivalently $e$) in (\ref{xpnfrB}).
The zeroth order is just $B^{(0)}=f$, while
\beqa B^{(1)}&=& -(f^2\theta +f\theta f + \theta f^2) +\frac 12 f
\;{\rm Tr}f\theta +\frac 14 \theta\; {\rm Tr}f^2\cr & &\cr
 B^{(2)}&=& \theta f^3 \theta + f\theta f^2\theta + \theta f^2
 \theta f + f (f \theta)^2 +(\theta f)^2 f + (f\theta)^2
 f \cr & &\cr & & - \frac 12(f^2\theta +f\theta f + \theta f^2)\; {\rm
   Tr}f\theta -\frac 14 \theta f \theta\;{\rm Tr} f^2 \cr & &\cr & & -\frac 12 \theta \;{\rm Tr}f^3\theta +\frac 18 \theta\;{\rm Tr} f\theta \;{\rm Tr} f^2+ \frac 18  f\;({\rm Tr}f\theta)^2 -\frac 14 f\; {\rm Tr}(f\theta)^2 
\eeqa

ii) Up to the first class constraints, the Hamiltonian density in (\ref{tlnrgdns}) can be re-expressed as
\be {\cal H}=2\frac{\partial {\cal L}}{\partial f_{0i}}\;f_{0i}  -
{\cal L} \ee  The coefficients ${\cal H}^{(n)}$ in the expansion given in (\ref{tlnrgdns}) are then given by
 \be {\cal H}^{(n)}= -(fB^{(n)})_{00}  -
{\cal L}^{(n)}\;,
\ee and so
\beqa  {\cal H}^{(0)}&=& \frac 12 (f_{i0})^2 + \frac 14 (f_{ij})^2\cr
& &\cr {\cal H}^{(1)}&=&(f^3\theta + f^2 \theta f + f\theta f^2)_{00}+\frac 12{\rm Tr}f^3\theta+ \frac 12  {\cal
  H}^{(0)}{\rm Tr}f\theta -\frac 14 (f\theta)_{00}{\rm Tr}f^2\cr
& &\cr {\cal H}^{(2)}&=& -\Bigl((f^2\theta)^2 + f\theta f^2 \theta f + f^2 ( f \theta  )^2 + f^2 (\theta f)^2 
+(f\theta)^2f^2 + f\theta f^3\theta \Bigr)_{00}\cr & &\cr & &- \frac 14\; {\rm Tr}( f^{2}\theta)^2-\frac
12 \;{\rm Tr} f^{2} (f\theta)^2 + \frac 12 (f\theta)_{00}\;  {\rm Tr} f^3\theta + \frac 14[(f\theta)^2]_{00} {\rm Tr} f^2 \cr & &\cr & & + \frac 12 {\cal H}^{(1)}\;{\rm Tr}f\theta- \frac 18 {\cal H}^{(0)}\;\Bigl(  ({\rm Tr}f\theta)^2+ 2{\rm Tr}(f\theta)^2\Bigr)\label{Hzrontw}\eeqa
The Hamiltonian function is properly expressed in terms of phase space variables, but as our purpose is only to use these expressions to examine the self-energy  for the solutions given in section 4, it is not necessary to re-express ${\cal H}$ in terms of such variables.


\bigskip

\begin{thebibliography}{99}


\bibitem{Douglas:2001ba}
  M.~R.~Douglas and N.~A.~Nekrasov,
  Rev.\ Mod.\ Phys.\  {\bf 73}, 977 (2001).

\bibitem{Szabo:2001kg}
  R.~J.~Szabo,
  Phys.\ Rept.\  {\bf 378}, 207 (2003).

\bibitem{Grosse:2004yu}
  H.~Grosse and R.~Wulkenhaar,  
  JHEP {\bf 0312}, 019 (2003);
  Commun.\ Math.\ Phys.\  {\bf 256}, 305 (2005).


  
 \bibitem{Blaschke:2008yj}
  D.~N.~Blaschke, F.~Gieres, E.~Kronberger, M.~Schweda and M.~Wohlgenannt,
  arXiv:0804.1914 [hep-th].

 \bibitem{Balachandran:2005pn}
  A.~P.~Balachandran, A.~Pinzul and B.~A.~Qureshi,
  Phys.\ Lett.\  B {\bf 634}, 434 (2006).
  
   \bibitem{Balachandran:2008gr}
  A.~P.~Balachandran, A.~Pinzul and A.~R.~Queiroz,
  arXiv:0804.3588 [hep-th].

 \bibitem{Steinacker:2007dq}
  H.~Steinacker,
  JHEP {\bf 0712}, 049 (2007);    
  arXiv:0712.3194 [hep-th].


 
 
 \bibitem{Grosse:2008xr}
  H.~Grosse, H.~Steinacker and M.~Wohlgenannt,
  JHEP {\bf 0804}, 023 (2008).

 
 
\bibitem{Hashimoto:1999zw}
  A.~Hashimoto and K.~Hashimoto,
  JHEP {\bf 9911}, 005 (1999).

\bibitem{Bak:1999id}
  D.~Bak,
  Phys.\ Lett.\  B {\bf 471}, 149 (1999).

\bibitem{Nekrasov:2000ih}
  N.~A.~Nekrasov,
  arXiv:hep-th/0011095.

\bibitem{Lozano:2000qf}
  G.~S.~Lozano, E.~F.~Moreno and F.~A.~Schaposnik,
  Phys.\ Lett.\  B {\bf 504}, 117 (2001).


\bibitem{Martin:2005vr}
  C.~P.~Martin and C.~Tamarit,
  JHEP {\bf 0602}, 066 (2006);   {\bf 0701}, 100 (2007).

\bibitem{Stern:2006zt}
  A.~Stern,
  Nucl.\ Phys.\  B {\bf 745}, 236 (2006).


\bibitem{Chamseddine:2000si}
  A.~H.~Chamseddine,
  Phys.\ Lett.\  B {\bf 504}, 33 (2001).

\bibitem{Aschieri:2005yw}
  P.~Aschieri, C.~Blohmann, M.~Dimitrijevic, F.~Meyer, P.~Schupp and J.~Wess,
  Class.\ Quant.\ Grav.\  {\bf 22}, 3511 (2005).

\bibitem{Calmet:2006iz}
  X.~Calmet and A.~Kobakhidze,
  Phys.\ Rev.\  D {\bf 74}, 047702 (2006).
  
\bibitem{Banerjee:2007th}
  R.~Banerjee, P.~Mukherjee and S.~Samanta,
  Phys.\ Rev.\  D {\bf 75}, 125020 (2007).
  


\bibitem{Nicolini:2005vd}
  P.~Nicolini, A.~Smailagic and E.~Spallucci,
  Phys.\ Lett.\  B {\bf 632}, 547 (2006).



\bibitem{Chaichian:2007we}
  M.~Chaichian, A.~Tureanu and G.~Zet,
  Phys.\ Lett.\  B {\bf 660}, 573 (2008).

\bibitem{Mukherjee:2007fa}
  P.~Mukherjee and A.~Saha,
  Phys.\ Rev.\  D {\bf 77}, 064014 (2008).

\bibitem{Kobakhidze:2007jn}
  A.~Kobakhidze,
  arXiv:0712.0642 [gr-qc].

\bibitem{Pinzul:2005ta}
  A.~Pinzul and A.~Stern,
  Class.\ Quant.\ Grav.\  {\bf 23}, 1009 (2006).

 
\bibitem{Smailagic:2003rp}
  A.~Smailagic and E.~Spallucci,
  J.\ Phys.\ A  {\bf 36}, L467 , L517 (2003).
  
\bibitem{bi} M. Born and L. Infeld, Proc. Roy. Soc. London {\bf A144} 425 (1934).


\bibitem{Seiberg:1999vs}
N.~Seiberg and E.~Witten,
JHEP {\bf 9909}, 032 (1999).


\bibitem{groe} H. Groenewold, Physica (Amsterdam) {\bf 12}, 405 (1946).

\bibitem{moy} J. Moyal, Proc. Camb. Phil. Soc. {\bf 45}, 99 (1949).

\bibitem{Gaete:2003dh}
  P.~Gaete and I.~Schmidt,
  Int.\ J.\ Mod.\ Phys.\  A {\bf 19}, 3427 (2004).
  
  
  \bibitem{Stern:2007an}
  A.~Stern,
  Phys.\ Rev.\ Lett.\  {\bf 100}, 061601 (2008).

\bibitem{Balachandran:2004rq}
  A.~P.~Balachandran, T.~R.~Govindarajan, C.~Molina and P.~Teotonio-Sobrinho,
  JHEP {\bf 0410}, 072 (2004).


\bibitem{Balachandran:2004cr}
  A.~P.~Balachandran and A.~Pinzul,
  Mod.\ Phys.\ Lett.\  A {\bf 20}, 2023 (2005).
  


\bibitem{Jurco:2001rq}
  B.~Jurco, L.~Moller, S.~Schraml, P.~Schupp and J.~Wess,
  Eur.\ Phys.\ J.\  C {\bf 21}, 383 (2001).
  
  \bibitem{Goto:2000zj}
  S.~Goto and H.~Hata,
  Phys.\ Rev.\  D {\bf 62}, 085022 (2000).

\bibitem{Brace:2001fj}
  D.~Brace, B.~L.~Cerchiai, A.~F.~Pasqua, U.~Varadarajan and B.~Zumino,
  JHEP {\bf 0106}, 047 (2001).
  
\bibitem{Fidanza:2001qm}
  S.~Fidanza,
  JHEP {\bf 0206}, 016 (2002).

\bibitem{Moller:2004qq}
  L.~Moller,
  JHEP {\bf 0410}, 063 (2004).


\bibitem{Trampetic:2007hx}
  J.~Trampetic and M.~Wohlgenannt,
  Phys.\ Rev.\  D {\bf 76}, 127703 (2007).

\bibitem{Gomis:2000sp}
  J.~Gomis, K.~Kamimura and T.~Mateos,
  JHEP {\bf 0103}, 010 (2001).

\bibitem{Bichl:2001gu}
  A.~A.~Bichl, J.~M.~Grimstrup, L.~Popp, M.~Schweda and R.~Wulkenhaar,
  arXiv:hep-th/0102103.


\bibitem{Banerjee:2003vc}
  R.~Banerjee, C.~k.~Lee and H.~S.~Yang,
  Phys.\ Rev.\  D {\bf 70}, 065015 (2004).


\bibitem{Banerjee:2005yy}
  R.~Banerjee and K.~Kumar,
  Phys.\ Rev.\  D {\bf 71}, 045013 (2005).

 



\bibitem{Eides:2000xc}
  M.~I.~Eides, H.~Grotch and V.~A.~Shelyuto,
  Phys.\ Rept.\  {\bf 342}, 63 (2001).
  \bibitem{Ho:2001aa}
  P.~M.~Ho and H.~C.~Kao,
  Phys.\ Rev.\ Lett.\  {\bf 88}, 151602 (2002). 

  
\bibitem{Chaichian:2000si}
  M.~Chaichian, M.~M.~Sheikh-Jabbari and A.~Tureanu,
  Phys.\ Rev.\ Lett.\  {\bf 86}, 2716 (2001);
  Eur.\ Phys.\ J.\  C {\bf 36}, 251 (2004).



\bibitem{OBJGRCMLDZ}
O.~Bertolami, J.~ G.~Rosa, C.~M.~L.~de~Arag\~ao, P.~Castorina, and D.~Zappal\`a,
Phys.\ Rev.\ D {\bf 72}, 025010 (2005).



\bibitem{OBJGR}
O.~Bertolami, J.~ G.~Rosa, 
  J.\ Phys.: Conf.\ Ser.  {\bf 33}, 118 (2006).


\bibitem{RBBDRSS}
R.~Banerjee, B.~Dutta~Roy, S.~Samanta, 
Phys.\ Rev.\ D {\bf 74}, 045015 (2006). 

\bibitem{Saha:2006xx}
  A.~Saha,
  Eur.\ Phys.\ J.\  C {\bf 51}, 199 (2007).



\bibitem{vor} F. Bayen, in {\it Group Theoretical Methods in Physics},
ed. E. Beiglb\"ock ,
et. al. [Lect. Notes Phys. {\bf 94}, 260 (1979)]; A. Voros, Phys. Rev. {\bf A 40},
6814 (1989).

\bibitem{Alexanian:2000uz}
  G.~Alexanian, A.~Pinzul and A.~Stern,
  Nucl.\ Phys.\  B {\bf 600}, 531 (2001).


\bibitem{Kontsevich:1997vb}
  M.~Kontsevich,
  Lett.\ Math.\ Phys.\  {\bf 66} (2003) 157.

\bibitem{Pinzul:2007bk}
  A.~Pinzul and A.~Stern,
  Nucl.\ Phys.\  B {\bf 791}, 284 (2008).


\bibitem{Pinzul:2003vp}
  A.~Pinzul and A.~Stern,
  Nucl.\ Phys.\  B {\bf 676}, 325 (2004).

\end{thebibliography}
\end{document}